\documentclass{article}

\usepackage[preprint]{neurips_2025}

\usepackage[utf8]{inputenc}
\usepackage[T1]{fontenc}
\usepackage{hyperref}
\usepackage{url}
\usepackage{booktabs}
\usepackage{amsmath}
\usepackage{amsfonts}
\usepackage{nicefrac}
\usepackage{microtype}
\usepackage{xcolor}
\usepackage{graphicx}
\usepackage{tikz}
\usetikzlibrary{positioning, arrows.meta, shapes.geometric}

\hypersetup{
  colorlinks=true,
  linkcolor=blue!50!black,
  citecolor=blue!50!black,
  urlcolor=blue!60!black,
}

\title{LogDx-CI: Benchmarking Log Reduction Tools for\\ LLM Root-Cause Diagnosis}

\author{%
  Bowen Qin \\
  National University of Singapore \\
  \href{https://github.com/eyuansu62/LogDx}{\texttt{github.com/eyuansu62/LogDx}}
}

\begin{document}

\maketitle

\begin{figure}[!h]
\centering
\includegraphics[width=0.88\linewidth]{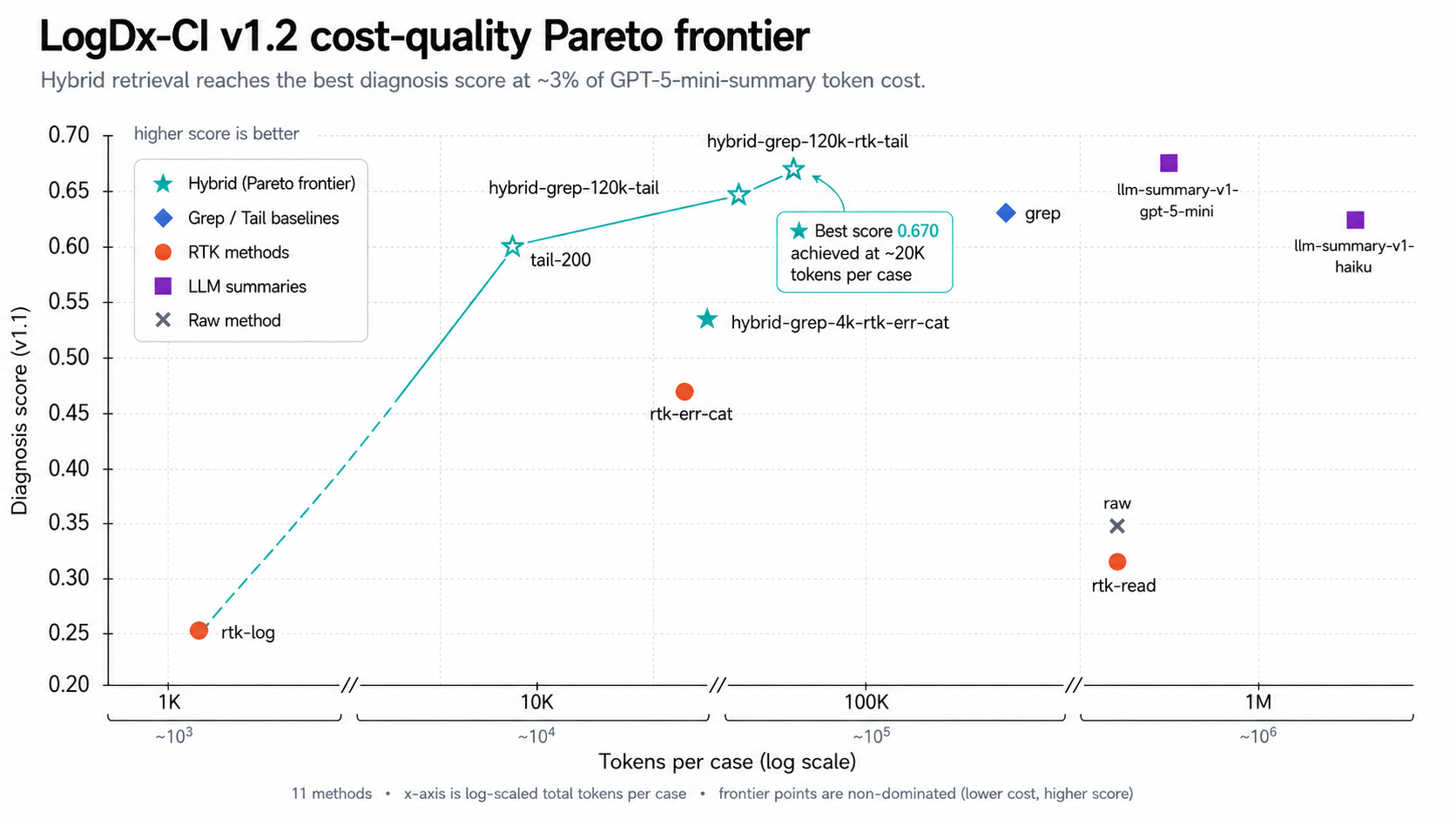}
\caption{\textbf{Cost-quality Pareto frontier across 11
  log-reduction tools.} Case-count-weighted macro
  \texttt{diagnosis\_score\_v1\_1} across the 35-case corpus
  $\times$ 3 single-shot LLM debugger families. Green dashed line
  traces the 4-method frontier
  (\texttt{rtk-log} $\to$ \texttt{tail-200} $\to$
  \texttt{hybrid-grep-120k-tail} $\to$
  \texttt{hybrid-grep-120k-rtk-tail}); all other methods are
  dominated. The top two hybrids reach 0.670 / 0.666 at $\sim$20k
  tokens per case --- \textbf{4.5$\times$ fewer tokens than
  \texttt{grep} at same-ballpark quality}. See
  \S\ref{sec:headline}--\S\ref{sec:agentloop}.}
\label{fig:pareto}
\end{figure}

\begin{abstract}
CI failure logs are large (median 5k lines, max 200k in this corpus) and
noisy. Coding agents that try to debug them depend on an upstream tool
to reduce the log to a manageable context, but the field has had no
public empirical comparison of which reductions preserve enough evidence
for downstream LLM diagnosis. We introduce \textbf{LogDx-CI}, a benchmark
that compares \textbf{11 context-reduction tools} (raw, tail, grep, three
RTK modes, two real LLM map-reduce summarizers, three hybrid routers) on
\textbf{35 real GitHub Actions failure cases}, scored by \textbf{3 LLM
debugger families} (Claude Haiku 4.5, Claude Sonnet 4.6, OpenAI
gpt-5-mini) plus a Sonnet 4.6 tool-using agent. We report three
load-bearing findings. (1)~Hybrid grep+tail routers dominate the
cost-quality Pareto frontier; the top two methods score 0.670 / 0.666
at $\sim$\$0.03 per case, same-ballpark quality as standalone grep at
$4.5\times$ fewer tokens. (2)~In the agent-loop regime, the quality
range across reduction tools collapses $7\times$ (single-shot spread
0.42 $\to$ agent-loop spread 0.059); the agent rescues weak contexts
via follow-up tool calls. However, cost differences persist: weak
contexts force the agent to issue 2--4$\times$ more tool calls to
recover. (3)~A cross-family LLM-summary pair (gpt-5-mini summarizer
feeding a Claude Haiku debugger) beats the same-family pair by
$+0.071$ averaged across four diagnoser variants, falsifying the
self-call-bias hypothesis on this task. The gpt-5-mini summarizer is
also the agent-loop \#1 method (score 0.749) at $0.37$ tool-calls per
case and $10\times$ lower reducer cost than the Haiku summarizer
(\$0.18 vs \$1.75 per case). All data, code, per-case bundles, and
reproducibility infrastructure are public.
\end{abstract}

\section{Introduction}

LogDx-CI evaluates \textbf{whether a log reduction tool preserves
enough evidence for an LLM to identify the root cause of a CI
failure}. The pipeline is:

\begin{center}
\begin{tikzpicture}[
  node distance=8mm,
  every node/.style={font=\small},
  stage/.style={
    rectangle, rounded corners=2pt,
    draw=black!55, line width=0.5pt,
    fill=gray!6,
    minimum height=11mm, inner xsep=4mm, inner ysep=2mm,
    align=center,
  },
  arr/.style={-{Stealth[length=2mm,width=2mm]}, draw=black!60, line width=0.6pt},
]
  \node[stage] (raw)  {raw CI log};
  \node[stage, right=of raw]  (red)  {context\\reduction tool};
  \node[stage, right=of red]  (dbg)  {LLM diagnoser\\(or 5-turn agent)};
  \node[stage, right=of dbg]  (sco)  {\texttt{diagnosis\_}\\\texttt{score\_v1\_1}};

  \draw[arr] (raw) -- (red);
  \draw[arr] (red) -- (dbg);
  \draw[arr] (dbg) -- (sco);
\end{tikzpicture}
\end{center}

The corpus is real public GitHub Actions failures across 8 categories
and 7+ ecosystems. Ground truth is AI-drafted (Claude Opus 4.7) +
single-author verified.

\paragraph{Why this matters.} CI failure logs in this corpus range
from 27 lines to 200k+, with a median around 5k. Most exceed the
effective input window of even long-context models once tool
definitions, system prompts, and reasoning overhead are accounted
for. A reduction step is therefore nearly mandatory in production
agent stacks (Claude Code, Codex, Cursor all ship some form of it),
but the field has had no public empirical comparison of which
reductions preserve enough evidence for downstream LLM diagnosis.
This benchmark closes that gap.

\paragraph{Scope.} This is a \textbf{niche benchmark}, not a general
agent evaluation. It does not measure: general coding ability,
multi-step planning, repository navigation, or open-ended debugging.
Findings should not be extrapolated beyond ``single-LLM-call or
5-turn-agent diagnosis of a single CI failure log.''

\section{Related Work}

\paragraph{Coding-agent benchmarks.}
The dominant benchmark for LLM-based software engineering is
SWE-bench~\cite{jimenez2024swebench}, which measures whether an
agent can resolve real GitHub issues by editing a repository.
Terminal-Bench~\cite{terminalbench2024} measures agent task
completion in a terminal environment, a closer analog to our
CI-failure-diagnosis setting but optimizing for end-to-end task
success rather than diagnostic accuracy.
WebArena~\cite{zhou2024webarena} and OSWorld~\cite{xie2024osworld}
benchmark agents in web and OS environments respectively. None of
these benchmarks isolate the \textbf{context-reduction step} that
precedes LLM diagnosis---the upstream tool that selects what
evidence reaches the model is treated as opaque infrastructure. We
measure this step directly.

\paragraph{LLM-as-judge evaluation.}
Many recent benchmarks rely on LLMs to score outputs: MT-Bench and
Chatbot Arena~\cite{zheng2023judging},
AlpacaEval~\cite{li2023alpacaeval}, G-Eval~\cite{liu2023geval}, and
RAGAS~\cite{es2024ragas} for retrieval-augmented generation.
Several biases complicate this approach.
Zheng~et~al.~\cite{zheng2023judging} document
\textbf{self-enhancement bias} (models prefer their own outputs in
pairwise comparison) and \textbf{position bias} (the first response
wins in pairwise judging disproportionately often). Our
Section~\ref{sec:crossfamily} measurement extends this literature
with a different bias variant: whether a summarizer plus downstream
judge of the same vendor family produces inflated scores via shared
priors. We find this \textbf{family-self-call bias is not present}
on our task; cross-family pairs beat same-family by $+0.071$ on
average across diagnoser variants.

\paragraph{Log compression and parsing.}
Classical log-parsing work (Drain~\cite{he2017drain},
Spell~\cite{du2016spell}, LogPAI and LogHub~\cite{zhu2019tools})
optimizes log compression for \emph{storage and human search}. Our
work differs in the objective function: we measure how much of the
failure signal survives various log-reduction strategies as judged
by downstream LLM diagnostic ability, rather than measuring
compression ratio or human-search recall. The RTK~\cite{rtk} tool,
which is a primary baseline in our benchmark, is an open-source
context-reduction CLI without a published benchmark of its
CI-debugging effectiveness---this benchmark is, in part, that
missing measurement.

\paragraph{Context selection for LLMs.}
A growing literature studies how LLM performance degrades with
context size and content structure: Lost-in-the-Middle
\cite{liu2024lostinthemiddle} shows that LLMs underweight
information in the middle of long inputs. Self-RAG~\cite{asai2024selfrag}
adds reflection to retrieval-augmented generation. Our hybrid
routers---the top-2 baselines on this benchmark---are a particularly
simple instance of this design space: a 120k-token threshold rule
that empirically identifies the abstain cliff for Sonnet 4.6 and
Haiku 4.5 on this corpus, with a deterministic fallback to
\texttt{tail-200}. The $7\times$ quality-range collapse in
agent-loop (Section~\ref{sec:agentloop}) is consistent with recent
findings that tool-use rescues weak retrieval in agent settings.

\section{Methodology}

\subsection{Corpus}

35 real GitHub Actions failure cases across 6 splits (Table~\ref{tab:corpus}):
8 failure categories
(\texttt{test\_assertion}, \texttt{compile\_error}, \texttt{type\_error},
\texttt{lint\_failure}, \texttt{dependency\_install},
\texttt{docker\_build}, \texttt{timeout\_or\_oom},
\texttt{multi\_failure}), 7+ ecosystems (pytest, cargo,
\texttt{go~test}, Maven, pnpm+jest, docker buildx, helm/k8s, etc.),
and log sizes from 27 lines to 200k+.

\begin{table}[h]
\centering
\caption{Corpus splits.}
\label{tab:corpus}
\begin{tabular}{lrl}
\toprule
\textbf{Split} & \textbf{Cases} & \textbf{Wave} \\
\midrule
\texttt{dev}        & 5  & v1 (prototype) \\
\texttt{holdout}    & 5  & v1 (prototype) \\
\texttt{stress}     & 6  & v1 (prototype) \\
\texttt{v2/dev}     & 3  & v2 (formal) \\
\texttt{v2/holdout} & 10 & v2 (formal) \\
\texttt{v2/stress}  & 6  & v2 (formal) \\
\midrule
\textbf{Total}      & \textbf{35} & \\
\bottomrule
\end{tabular}
\end{table}

Each case carries five files under
\texttt{cases/<split>/<case\_id>/}: \texttt{raw.log},
\texttt{case.json} (metadata), \texttt{ground\_truth.json} (root
cause, required signals, relevant files/tests, must-mention
checklist, forbidden claims), \texttt{tags.json}, and
\texttt{privacy\_audit.json}.

Logs are sourced from publicly visible GitHub Actions runs. Each
passed through a privacy audit (200k-line cap, fail-closed on
truncation, URL / bearer / API-key / long-opaque-token redactors)
before commit. Zero redaction hits across all 35 cases.

\subsection{Evaluation metric}

The primary metric \texttt{diagnosis\_score\_v1\_1} is a calibrated
linear combination of category accuracy, required-signal recall,
relevant-file and relevant-test recall, must-mention coverage from
the ground-truth checklist, and valid-evidence-quote rate.
Penalized for forbidden claims and confident errors (confidence
$\geq 0.70$ and demonstrably wrong on multiple fronts---the metric
closest to ``this method led the LLM to confidently misdiagnose'').

We also report \texttt{confident\_error\_rate\_v1\_1} as a separate
column because confidently-wrong diagnoses are operationally
distinct from ``missed the diagnosis''; the safety implications
differ.

\subsection{Baselines --- 11 context providers}
\label{sec:baselines}

\begin{table}[h]
\centering
\small
\caption{11 baseline context providers.}
\label{tab:baselines}
\begin{tabular}{ll}
\toprule
\textbf{Provider} & \textbf{Implementation} \\
\midrule
\texttt{raw} & Full log handed to the model \\
\texttt{tail-200} & Last 200 lines \\
\texttt{grep} & Regex-filtered failure-pattern lines + 3/8 context \\
\texttt{rtk-read} & RTK in read mode \\
\texttt{rtk-log} & RTK in log mode \\
\texttt{rtk-err-cat} & RTK in error-category mode \\
\texttt{llm-summary-v1-haiku}     & Real Haiku 4.5 map-reduce summarizer$^{\dagger}$ \\
\texttt{llm-summary-v1-gpt-5-mini} & Real gpt-5-mini map-reduce summarizer$^{\dagger}$ \\
\texttt{hybrid-grep-4k-rtk-err-cat} & Earlier 4k-threshold hybrid (replaced) \\
\texttt{hybrid-grep-120k-tail} & grep $\leq$ 120k tokens else tail-200 \\
\texttt{hybrid-grep-120k-rtk-tail} & grep $\leq$ 120k else rtk-err-cat (if not truncated) else tail-200 \\
\bottomrule
\end{tabular}
\end{table}

The 120k threshold is empirically the abstain cliff for Sonnet 4.6
and Haiku 4.5 on this corpus; above it, the model context window
plus diagnostic prompt overhead causes abstention. See
Section~\ref{sec:failuremodes} for the density-driven inflation
failure mode that motivates the choice.

$^{\dagger}$ All map-reduce summarizers use \texttt{chunk\_lines=500},
\texttt{chunk\_overlap\_lines=25}, \texttt{temperature=0}. Three of
the 35 cases used \texttt{chunk\_lines=100} instead for the Haiku
summarizer because they contained 500-line windows exceeding
Haiku's effective input window after Claude-Code session overhead;
recorded in per-case \texttt{metadata.chunk\_lines}.

\subsection{Diagnosers}

Four diagnosers receive the same prompt template and produce the
same diagnosis JSON schema. Outputs are evaluated by the same
deterministic evaluator.

\begin{itemize}
  \item \texttt{real-debugger-v1}: Claude Haiku 4.5, single-shot.
  \item \texttt{real-debugger-v2}: Claude Sonnet 4.6, single-shot.
  \item \texttt{real-debugger-v3}: OpenAI gpt-5-mini (pinned to \texttt{gpt-5-mini-2025-08-07}), single-shot.
  \item \texttt{real-agent-v1}: Claude Sonnet 4.6 plus four
        deterministic tools (\texttt{grep}, \texttt{read\_file},
        \texttt{tail}, \texttt{view\_log\_lines}) operating on the
        raw log. 5-turn cap, 180k cumulative-input cap.
\end{itemize}

The agent variant operates on the \emph{raw log}: tool calls bypass
the upstream context provider and let the agent re-query the
underlying log when the initial reduction is insufficient. This
separates ``context quality'' from ``agent recovery capability.''

\section{Results}

\subsection{Single-shot headline}
\label{sec:headline}

Table~\ref{tab:singleshot} reports \texttt{diagnosis\_score\_v1\_1}
case-count-weighted macro across the 35-case corpus, aggregated over
the three single-shot debugger families.

\begin{table}[h]
\centering
\small
\caption{Single-shot leaderboard. \texttt{conf\_err} is the rate of
  confidently-wrong diagnoses (lower is better).}
\label{tab:singleshot}
\resizebox{\linewidth}{!}{%
\begin{tabular}{rlrrrrrr}
\toprule
\textbf{Rank} & \textbf{Method} & \textbf{Haiku} & \textbf{Sonnet} & \textbf{gpt-5-mini} & \textbf{Overall} & \textbf{conf\_err} & \textbf{tokens/case} \\
\midrule
1 & \texttt{hybrid-grep-120k-rtk-tail}  & 0.624 & 0.679 & 0.706 & \textbf{0.670} & 0.000 & 19{,}844 \\
2 & \texttt{hybrid-grep-120k-tail}      & 0.610 & 0.730 & 0.658 & \textbf{0.666} & 0.010 & 19{,}753 \\
3 & \texttt{llm-summary-v1-gpt-5-mini}  & 0.654 & 0.686 & 0.652 & \textbf{0.664} & 0.010 & 537{,}638 \\
4 & \texttt{grep}                       & 0.578 & 0.684 & 0.655 & 0.639 & 0.000 & 88{,}355 \\
5 & \texttt{llm-summary-v1-haiku}       & 0.583 & 0.704 & 0.608 & 0.632 & 0.029 & 1{,}681{,}520 \\
6 & \texttt{tail-200}                   & 0.595 & 0.624 & 0.623 & 0.614 & 0.019 & 6{,}108 \\
7 & \texttt{hybrid-grep-4k-rtk-err-cat} & 0.552 & 0.597 & 0.571 & 0.573 & 0.029 & 19{,}892 \\
8 & \texttt{rtk-err-cat}                & 0.455 & 0.488 & 0.467 & 0.470 & 0.029 & 19{,}850 \\
9 & \texttt{raw}                        & 0.324 & 0.368 & 0.367 & 0.353 & 0.000 & 275{,}248 \\
10 & \texttt{rtk-read}                  & 0.329 & 0.369 & 0.349 & 0.349 & 0.010 & 274{,}289 \\
11 & \texttt{rtk-log}                   & 0.238 & 0.262 & 0.249 & 0.249 & 0.133 & 810 \\
\bottomrule
\end{tabular}%
}
\end{table}

Three findings from Table~\ref{tab:singleshot}:
(i)~Both Pareto-winning hybrids beat every single-method baseline
on quality. The $+0.03$ to $+0.04$ gap over \texttt{grep} is modest,
but the token-cost axis is $4.5\times$ cheaper.
(ii)~The top-3 methods produce zero or near-zero confidently-wrong
diagnoses. \texttt{rtk-log} misleads a confident LLM on
$\sim$13\,\% of cases.
(iii)~\texttt{grep} is dominated by \texttt{hybrid-grep-120k-tail}
on both axes: same-ballpark score (0.639 vs.\ 0.666) at
$4.5\times$ fewer tokens (88{,}355 vs.\ 19{,}753).

\subsection{Cross-debugger stability}

Table~\ref{tab:topthree} shows the top-3 method set under each
single-shot debugger family.

\begin{table}[h]
\centering
\small
\caption{Top-3 sets per debugger family.}
\label{tab:topthree}
\begin{tabular}{ll}
\toprule
\textbf{Family} & \textbf{Top-3} \\
\midrule
Claude Haiku 4.5  & \texttt{hybrid-grep-120k-rtk-tail}, \texttt{hybrid-grep-120k-tail}, \texttt{tail-200} \\
Claude Sonnet 4.6 & \texttt{hybrid-grep-120k-tail}, \texttt{grep}, \texttt{hybrid-grep-120k-rtk-tail} \\
OpenAI gpt-5-mini & \texttt{hybrid-grep-120k-rtk-tail}, \texttt{hybrid-grep-120k-tail}, \texttt{grep} \\
\bottomrule
\end{tabular}
\end{table}

The \textbf{top-3 intersection} across all three families is
\{\texttt{hybrid-grep-120k-rtk-tail}, \texttt{hybrid-grep-120k-tail}\}
---both 120k-threshold hybrids. The \textbf{bottom-4 set} is also
stable across all three families: \{\texttt{raw}, \texttt{rtk-read},
\texttt{rtk-log}, \texttt{rtk-err-cat}\}. The cross-family agreement
is a direct robustness check; the direction is preserved across two
vendors (Anthropic + OpenAI) and two within-Anthropic capability
tiers (Haiku + Sonnet).

\subsection{Cost-quality Pareto frontier}
\label{sec:pareto}

Figure~\ref{fig:pareto} (on page~1) visualizes the cost-quality
trade-off. The 4-method Pareto frontier is
\texttt{rtk-log} (810 tokens, 0.249) $\to$
\texttt{tail-200} (6{,}108, 0.614) $\to$
\texttt{hybrid-grep-120k-tail} (19{,}753, 0.666) $\to$
\texttt{hybrid-grep-120k-rtk-tail} (19{,}844, 0.670). Every other
method is dominated. Notable dominations:
\texttt{grep} is dominated by \texttt{hybrid-grep-120k-tail} at
$4.5\times$ fewer tokens; \texttt{rtk-err-cat} is dominated by
\texttt{hybrid-grep-120k-tail} at similar token cost but $+0.20$
higher score; and \texttt{llm-summary-v1-haiku} is the most
expensive method on the board at 1.68M tokens per case
(predominantly Claude-Code-CLI nested cached-prefix overhead) but
scores rank 5.

Pinned USD costs (per case, snapshot 2026-05-20): top-2 hybrids
\$0.031, \texttt{tail-200} \$0.012, \texttt{grep} \$0.129,
\texttt{llm-summary-v1-gpt-5-mini} \$0.184, \texttt{raw} \$0.392,
\texttt{llm-summary-v1-haiku} \$1.760.

\subsection{Agent-loop measurement}
\label{sec:agentloop}

The single-shot leaderboard tests
\texttt{log $\to$ reducer $\to$ single LLM call $\to$ answer}. Real
Claude Code / Codex usage looks different: the model can call
follow-up tools when its initial context is missing something. We
add the agent-loop measurement using \texttt{real-agent-v1} (Sonnet
4.6, 5-turn cap, 4 deterministic tools).

\begin{table}[h]
\centering
\small
\caption{Agent-loop leaderboard.}
\label{tab:agentloop}
\begin{tabular}{rlrrrrr}
\toprule
\textbf{Rank} & \textbf{Method} & \textbf{Single-shot} & \textbf{Agent} & $\Delta$ & \textbf{conf\_err} & \textbf{Tools/case} \\
\midrule
1 & \texttt{llm-summary-v1-gpt-5-mini} & 0.664 & \textbf{0.749} & +0.085 & 0.000 & 0.37 \\
2 & \texttt{hybrid-grep-120k-rtk-tail} & 0.670 & 0.747 & +0.077 & 0.000 & 0.97 \\
3 & \texttt{hybrid-grep-4k-rtk-err-cat}& 0.573 & 0.737 & +0.164 & 0.000 & 1.40 \\
4 & \texttt{hybrid-grep-120k-tail}     & 0.666 & 0.735 & +0.069 & 0.000 & 1.00 \\
5 & \texttt{rtk-read}                  & 0.349 & 0.735 & +0.386 & 0.000 & 1.46 \\
6 & \texttt{grep}                      & 0.639 & 0.722 & +0.083 & 0.029 & 1.20 \\
7 & \texttt{tail-200}                  & 0.614 & 0.710 & +0.096 & 0.029 & 0.69 \\
8 & \texttt{rtk-err-cat}               & 0.470 & 0.708 & +0.238 & 0.000 & 1.66 \\
9 & \texttt{llm-summary-v1-haiku}      & 0.632 & 0.690 & +0.058 & 0.057 & 0.71 \\
10 & \texttt{rtk-log}                  & 0.249 & 0.689 & +0.440 & 0.057 & 2.60 \\
11 & \texttt{raw}                      & 0.353 & 0.688 & +0.335 & 0.029 & 1.68 \\
\bottomrule
\end{tabular}
\end{table}

Findings:
(i)~The quality range collapses $7\times$. Single-shot spread is
0.42 (0.670 $-$ 0.249); agent-loop spread is 0.059 (0.749 $-$
0.690). The agent rescues weak contexts via tool calls;
\texttt{rtk-log} gains $+0.440$, \texttt{rtk-read} gains $+0.386$,
\texttt{raw} gains $+0.335$.
(ii)~Safety mostly collapses. Five of eleven methods sit at 0\,\%
confident-error in agent-loop.
(iii)~The top single-shot method holds in agent-loop:
\texttt{hybrid-grep-120k-rtk-tail} is agent-loop \#2 at 0.747, just
behind \texttt{llm-summary-v1-gpt-5-mini} at 0.749 (within Sonnet
temperature-0 variance), using only 0.97 tool calls per case.
(iv)~\texttt{llm-summary-v1-gpt-5-mini} uses the lowest tool count
of any method (0.37 / case --- half as many as
\texttt{tail-200}'s 0.69). The real gpt-5-mini summary front-loads
the failure signal so completely that the agent commits to a
diagnosis on turn 1 about 60\,\% of the time.

\subsection{Cross-family LLM-summary}
\label{sec:crossfamily}

A reviewer raised after v1.1: was the haiku-summary's headline
promotion anchored on Claude-family priors? Specifically, does the
self-call pair (Haiku-summarizer $\to$ Haiku-debugger) carry shared
prior bias that inflates its score?

v1.2 backfills a non-Anthropic summarizer
(\texttt{llm-summary-v1-gpt-5-mini}---real OpenAI gpt-5-mini
map-reduce, same prompts / \texttt{chunk\_lines} / temperature as
the Haiku summarizer) across the full 35-case $\times$ 4-diagnoser
matrix (Table~\ref{tab:crossfamily}).

\begin{table}[h]
\centering
\small
\caption{Haiku summarizer vs.\ gpt-5-mini summarizer, paired with
  each diagnoser.}
\label{tab:crossfamily}
\begin{tabular}{lrrr}
\toprule
\textbf{Diagnoser} & \textbf{Haiku-sum} & \textbf{gpt5mini-sum} & $\Delta$ \\
\midrule
\texttt{real-debugger-v1} (Haiku 4.5)  & 0.583 & \textbf{0.654} & $+0.071$ \\
\texttt{real-debugger-v2} (Sonnet 4.6) & \textbf{0.704} & 0.686 & $-0.018$ \\
\texttt{real-debugger-v3} (gpt-5-mini) & 0.608 & \textbf{0.652} & $+0.044$ \\
\texttt{real-agent-v1} (Sonnet+tools)  & 0.690 & \textbf{0.749} & $+0.059$ \\
\bottomrule
\end{tabular}
\end{table}

Cross-family beats self-pair in 3 of 4 diagnosers. Specifically:
Haiku-summary is best on the Sonnet debugger (not on Haiku), and
gpt-5-mini-summary is best on the Haiku debugger (not on
gpt-5-mini). There is no clean ``summarizer-and-debugger of the
same family'' advantage. Summary quality is driven by the
summarizer's ability to extract failure signal at chunk
granularity, not by shared priors between summarizer and
downstream debugger model families. The self-call-bias hypothesis
is \textbf{falsified} for this task.

The cost gap matters too. The gpt-5-mini summarizer costs $10\times$
less per reducer call than haiku-summary (\$0.18 vs \$1.75 per case).
The gap is Claude-Code-CLI nested-invocation overhead (cached-prefix
tokens) that the OpenAI-direct call doesn't carry.

\subsection{Failure modes}
\label{sec:failuremodes}

Two reproducible failure modes worth naming:

\paragraph{Density-driven context inflation for \texttt{grep} / \texttt{rtk-err-cat}.}
Logs where \texttt{error}|\texttt{failed} markers appear in
test-progress noise cause \texttt{grep} / \texttt{rtk-err-cat}
outputs to exceed the model's effective reasoning budget. The rust
compiletest (31k lines $\to$ 161k tokens after grep) and a nodejs
timeout case (10k lines $\to$ 359k tokens) both push Sonnet and
Haiku into abstain. \texttt{tail-200} survives by being
content-blind and bounded; this motivates the 120k hybrid threshold.

\paragraph{RTK truncation drops bounded failure structure.}
\texttt{rtk-err-cat}'s aggressive compression strips assertion
diffs, snapshot diffs, and structured compiler error blocks that
the diagnoser needs. On 5 of 8 v2-formal cases where the legacy
4k-threshold hybrid routed to \texttt{rtk-err-cat}, \texttt{grep}
would have done strictly better. This is the v1.3-prototype
``selection-by-method'' overfit that the v1.2 120k-threshold
hybrids correct.

\section{Recommendations}

Three takeaways for practitioners deploying LLM-based CI debugging:

\begin{enumerate}
  \item \textbf{Single-LLM-call diagnosis (no tool use):} use
  \texttt{hybrid-grep-120k-rtk-tail}. Top-1 across all three model
  families, zero confident-error, \$0.03 per case, $4.5\times$
  cheaper than standalone \texttt{grep}.
  \item \textbf{Tool-using agents (Claude Code, Codex-style):}
  \texttt{llm-summary-v1-gpt-5-mini} is the v1.2 default. Agent-loop
  \#1 at 0.749, lowest tool-call count (0.37 / case), \$0.18 / case
  end-to-end. The hybrid above is a close second (0.747, 0.97 tools
  per case) and is appropriate when an extra LLM preprocessing call
  is not acceptable.
  \item \textbf{Avoid \texttt{rtk-log} standalone.} Its 13.3\,\%
  confident-error rate (single-shot) means it actively misleads
  downstream LLMs $\sim$1 in 8 cases.
\end{enumerate}

What this benchmark \emph{does not} settle: configured / tuned RTK
performance (v1.2 tests stock invocations only), generalization to
non-Anthropic / non-OpenAI model families, performance on log shapes
outside the 35-case distribution (notably pre-step CI runner
output, build-system streams, monorepo matrix jobs without a single
failing leaf).

\section{Caveats and limitations}

This is a \textbf{v1.2 preprint}. Headline findings are robust enough
to ship; per-case magnitudes are preliminary.

\begin{enumerate}
  \item \textbf{35 cases.} Per-case variance can shift macro means by
  $\pm 0.05$ with future corpus expansion. The direction of the
  top-3 $\cap$ finding is robust across debugger families; absolute
  magnitudes are preliminary.
  \item \textbf{Ground truth is AI-drafted (Claude Opus 4.7) +
  single-author verified.} Not independent human annotation. The
  full 35-case set has not been re-scored by an outside party.
  \item \textbf{Three model families tested} (Anthropic Haiku 4.5 +
  Sonnet 4.6; OpenAI gpt-5-mini; OpenRouter Sonnet 4.6 for the
  agent-loop diagnoser). Two unique vendors. Adding Gemini, Llama,
  or DeepSeek is the most-leveraged follow-up.
  \item \textbf{gpt-5-mini reproducibility caveat.} gpt-5-mini
  exhibits run-to-run variance even at \texttt{temperature=0}
  (reasoning models sample reasoning traces freshly each call).
  Practical effects: macro means in the leaderboard tables are
  stable to $\pm 0.02$ across re-runs; per-case byte-identical
  reproduction is not guaranteed. The aggregate ranking is robust;
  specific cell numbers may shift by $\pm 0.02$.
  \item \textbf{Agent-loop soft-cap.} The 5-turn, 180k-cumulative
  input cap is soft. Despite guards, 18 of 350 v1.1 rows landed
  above 180k (max 273{,}654). Costs reported reflect actual usage,
  not the nominal cap.
  \item \textbf{20 historical exclusions} documented in
  \path{configs/historical_provider_error_exclusions.json}
  appear as zero-score abstentions in the eval denominator. These
  correspond to transient model / CLI / API failures during the
  2026-04 / 05 prototype sweeps, removed by the 2026-05-15 cleanups.
  \item \textbf{Hybrid threshold is the variable, not the method
  shape.} This benchmark cannot say ``hybrid as a strategy is bad.''
  The v1.2 result says ``the 120k-token threshold tuned on this
  35-case corpus is the empirical abstain cliff for Sonnet 4.6 and
  Haiku 4.5; a fresh corpus might calibrate to a different
  threshold.''
\end{enumerate}

\section{Reproducibility}

Every release carries: (i)~a \textbf{protocol lock} (SHA-pins 10
schemas + 3 prompts + 4 evaluators + 10 baselines + 35 case files
at the release commit); (ii)~\textbf{3 release gates} that fail CI
when any committed artifact drifts; (iii)~a \textbf{165-test suite}
covering unit, integration, and end-to-end paths;
(iv)~\textbf{cache identity validation}
(\texttt{metadata.diagnoser\_config\_sha256} and
\texttt{metadata.shim\_sha256} on every fresh row); and
(v)~\textbf{secret redaction} via URL / bearer / API-key /
long-opaque-token / hostname redactors.

\paragraph{Reproducing a published number.}

\begin{verbatim}
git clone https://github.com/eyuansu62/LogDx.git
cd LogDx && git checkout v1.2
python3 tools/validate_protocol_lock.py \
    --protocol protocols/logdx-ci-v2-partial-2026-05-20.lock.json
python3 tools/validate_committed_diagnosis_provider_errors.py
python3 tools/validate_eval_manifest_consistency.py
python3 tools/validate_diagnosis_vs_context_consistency.py
python3 tools/evaluate_diagnosis.py --split v2/dev \
    --diagnoser real-debugger-v3
\end{verbatim}

The cases corpus is mirrored at
\href{https://huggingface.co/datasets/eyuansu71/logdx-ci}{huggingface.co/datasets/eyuansu71/logdx-ci}
with a flat-metadata viewer for HF Dataset Viewer browsing. For the
full per-case bundle (\texttt{raw.log} + \texttt{ground\_truth} +
\texttt{tags} + \texttt{privacy\_audit}), fetch via
\texttt{huggingface\_hub.snapshot\_download}.

\section{Conclusion}

LogDx-CI v1.2 is the first public benchmark of CI log reduction
tools optimized for downstream LLM diagnostic accuracy across model
families. The three load-bearing findings (hybrid Pareto dominance,
$7\times$ agent-loop quality-range collapse, falsified self-call
bias) carry concrete implications for production coding-agent
stacks: at \$0.03 per case for single-shot diagnosis and at \$0.18
per case with a cross-family LLM-summary prefilter for agent-loop
diagnosis, the cost of getting context-reduction right is at least
$10\times$ less than the cost of getting it wrong. Future work
includes corpus expansion (target 50+ cases), additional model
families (Gemini, Llama, DeepSeek), configured-RTK baselines, and
independent third-party re-scoring of the 35-case ground-truth set.

\section*{Acknowledgments}

LogDx-CI benchmarks third-party log-reduction tools alongside its
own baselines. The \texttt{rtk-read}, \texttt{rtk-log}, and
\texttt{rtk-err-cat} baselines are three different invocations of
RTK~\cite{rtk} by rtk-ai. CI failure logs are sourced from publicly
visible GitHub Actions runs. Diagnoses are produced by Claude
(Anthropic) and gpt-5-mini (OpenAI).

\bibliographystyle{plainnat}
\bibliography{references}

\end{document}